\documentclass[aps,prl,twocolumn,superscriptaddress,longbibliography]{revtex4-2}

\usepackage{amsmath,amssymb}
\usepackage{graphicx}
\usepackage{booktabs}
\usepackage{xcolor}
\usepackage{bm}
\usepackage{siunitx}
\usepackage{array}
\usepackage{hyperref}

\newcommand{\GMm}{\mathrm{\Gamma}_1^-}
\newcommand{\Aone}{\mathrm{A}_1^-}
\newcommand{\Atwo}{\mathrm{A}_2^-}

\newcommand{\NbOI}{NbOI$_2$}
\newcommand{\TaOI}{TaOI$_2$}
\providecommand{\angstrom}{\text{\AA}}
\renewcommand{\angstrom}{\text{\AA}}

\begin{document}

\title{Ferroelectricity and Peierls transition in MOI$_2$ (M=Ta, Nb): a first-principle analysis}

\author{Carmel Dansou}
\affiliation{Smart Ferroic Materials Center, Arkansas Materials Institute and Department of Physics, University of Arkansas, Fayetteville, Arkansas 72701, USA}

\author{Charles Paillard}
\affiliation{Smart Ferroic Materials Center, Arkansas Materials Institute and Department of Physics, University of Arkansas, Fayetteville, Arkansas 72701, USA}

\author{Laurent Bellaiche}
\affiliation{Smart Ferroic Materials Center, Arkansas Materials Institute and Department of Physics, University of Arkansas, Fayetteville, Arkansas 72701, USA}
\affiliation{Department of Materials Science and Engineering, Tel Aviv University, Ramat Aviv, Tel Aviv 6997801, Israel}

\date{\today}

\begin{abstract}
The van der Waals ferroelectric NbOI$_2$ hosts a polar distortion that is commonly attributed to the Peierls dimerization of its Nb chains. Here we show, using first-principles calculations and a Landau-type energy, that the polar instability is in fact a distinct order parameter that does not stem from the Peierls mode condensation and that three main distinct modes are important in the whole MOI$_2$ family. They are a polar mode $P$, an antipolar mode $\eta$, and a Peierls mode $\delta$, whose stabilities and couplings strongly depend on the nature of the $M$ metal element, as evidenced by the fact that the isostructural, TaOI$_2$ system is not polar while possessing Peirls distortions.
\end{abstract}

\maketitle

\emph{Introduction.---}
Two-dimensional van der Waals (vdW) ferroelectrics (FEs) combine switchable polarization with mechanical flexibility, weak interlayer interaction, and compatibility with heterostructure engineering, making them attractive for nonvolatile memories, photonics, and flexoelectric devices~\cite{jin2022rise,qi2021review}. Among them, the recent discovered layered niobium oxydihalides NbOX$_2$ (X = Cl, Br, I) have attracted intense attention for robust in-plane ferroelectricity, high second-harmonic
generation, and predicted photostriction and piezoelectricity \cite{jia_niobium_2019,ye_2023, ye_ferroelectric_2021, mohebpour_origin_2024,zhang_giant_2024,abdelwahab_giant_2022}.

\begin{figure*}[!ht]
	\centering
	\includegraphics[width=0.75\textwidth]{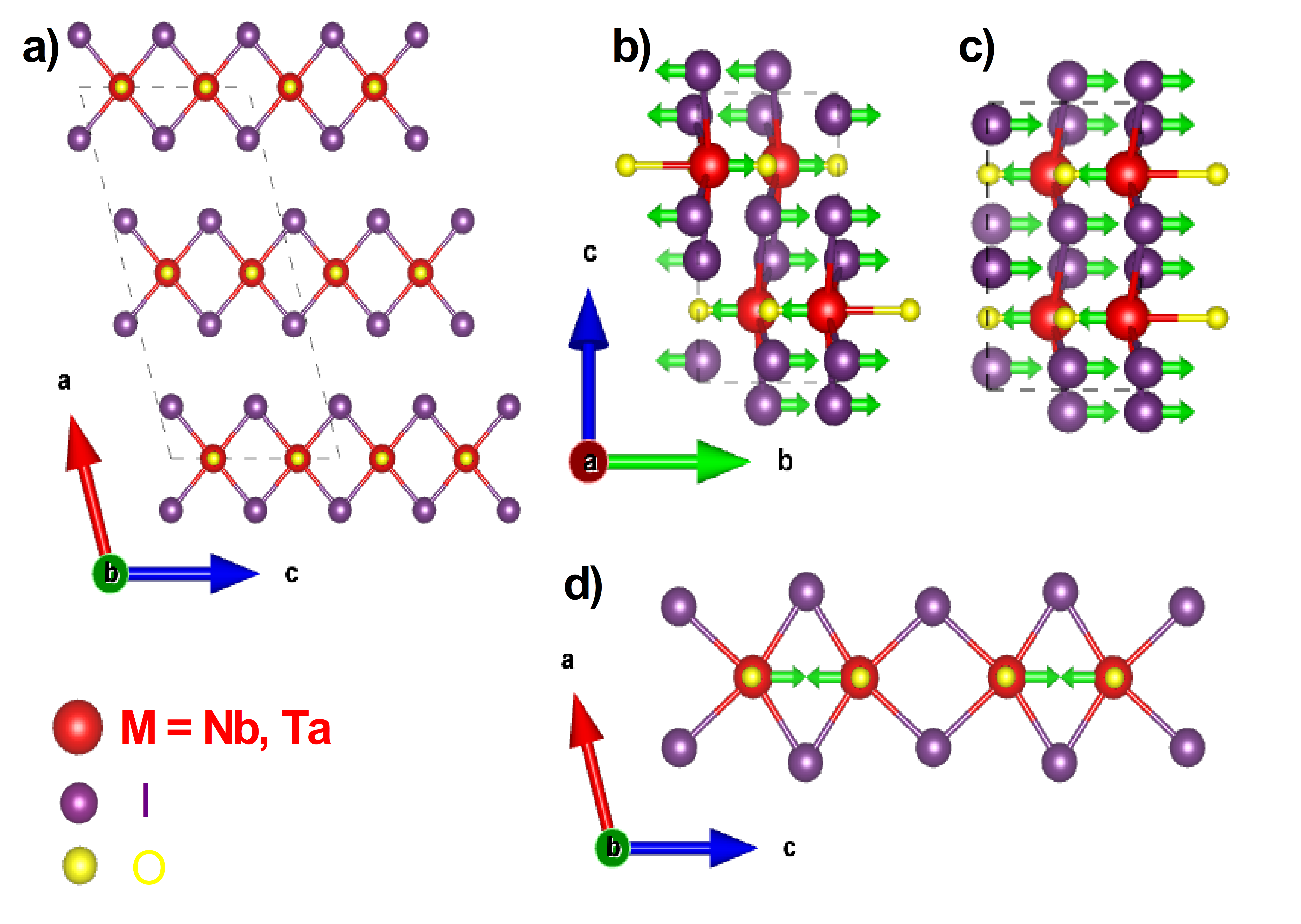}
	\caption{\label{f0} Crystal structure of MOI$_2$. (a) Crystal structure of the undimerized MOI$_2$ phase viewed along the $b$ axis. (b) Crystal structure of the antiferroelectric (AFE) phase viewed along the $a$ axis, with arrows indicating atomic displacements associated with the AFE mode. (c) Crystal structure of the ferroelectric (FE) phase viewed along the $a$ axis, with arrows indicating atomic displacements associated with the FE mode. (d) Crystal structure of the Peierls-distorted phase viewed along the $b$ axis, with arrows indicating atomic displacements associated with the Peierls distortion.}
\end{figure*}

\begin{figure}[!ht]
	\centering
	\includegraphics[width=0.5\textwidth]{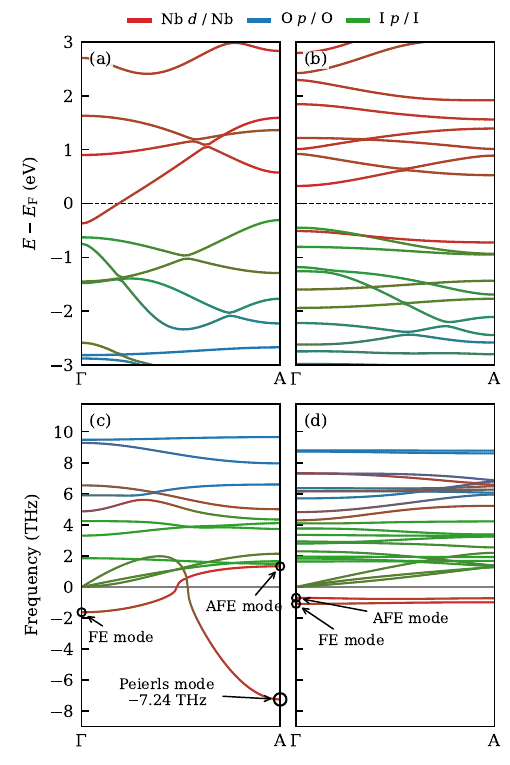}
	\caption{\label{f1}Electronic band structures (top) and phonon dispersions (bottom) of NbOI$_2$ along $\Gamma$-A for the ``4-atom cell" in panels (a), (c) and in the ``8-atom cell" (which is doubled along the Nb chain direction) in panels (b) and (d) t, . In (a) and (b), band colors show the orbital projections onto Nb d (red), O p (blue), and I p (green), and the dashed line marks the Fermi level. In (c) and (d), phonon branches are colored by the atomic contributions of Nb, O, and I to the eigenvector. The ``4-atom cell" is metallic, with a Nb-d band crossing E$_F$. It shows a FE instability at $\Gamma$ (-1.63 THz) and a strong Peierls instability at A (-7.24 THz), while the AFE mode at A is stable (1.33 THz). In the ``8-atom cell", a gap opens at E$_F$. The Peierls instability is removed, and the FE is hardened (-1.09 THz) and the AFE is softened (-0.70 THz). Both modes also now appear at $\Gamma$}
\end{figure}

Among the NbOX$_2$ allotrophes, NbOI$_2$ has been the most studied and was recently proposed as a host of ferronic physics \cite{choe2025, zhang2026electric}. The lattice of NbOI$_2$ contains one-dimensional Nb chains along the crystallographic $\mathbf{c}$ axis (see Fig\ref{f0}a), with a $d^1$ electron count per Nb. Along these chains, Nb--Nb dimerization is universally observed, and the standard narrative ties the material's ferroelectricity and some derived properties to this dimerization: the Peierls distortion of the half-filled chain opens a gap and, subsequently, is said to displace Nb off-center within its O$_2$I$_2$ cage, inducing the polar distortion along the $\mathbf{b}$ axis \cite{li2025peierls}. This Peierls-centric picture, however, remains incomplete in one important respect. The paraelectric (PE) reference phase of NbOI$_2$, usually used in the existing literature, exhibits \emph{two} soft modes at its Brillouin zone (BZ)-center (see phonon spectrum in Fig.\ref{f1}d): a FE mode in which the two Nb atoms of the cell displace in phase along the polar axis, and  an antiferroelectric (AFE) mode in which they displace in an antiphase fashion. First-principles calculations have shown that condensing the FE mode results into the observed ground state, and the AFE phase lies only $\sim$2~meV per formula unit higher \cite{jia_niobium_2019}. The reason why a Peierls' transition induces a \emph{pair} of polar and antipolar modes at $\Gamma$ is unexplained as AFE modes are at BZ boundary in conventional perovskites \cite{iniguez2014first, hlinka2014multiple}. Another key limitation is that previous studies have, to our knowledge, not gone beyond the already-dimerized eight-atom description to identify a smaller undimerized parent structure. By building the Peierls periodicity into the reference cell from the start, such a description obscures the translational-symmetry breaking associated with dimerization and, consequently, the microscopic origin of the instabilities and some physical properties of NbOI$_2$. Additionally, TaOI$_2$ is known to be nonpolar despite exhibiting an intrinsic Peierls distortion and a very similar electronic structure to NbOI$_2$ \cite{ruck1995taoi2,tang2024third, ding2025strain}. This is an interesting puzzle that the existing literature has failed to elucidate.

In this Letter, we answer these questions by showing that the conventional used PE phase is not the fundamental parent structure of NbOI$_2$. Using pseudosymmetry analysis, we construct a smaller one-formula-unit (i.e., four-atom) phase in which Nb occupies an inversion center, and of which the ``eight atoms" cell is itself a Peierls-condensed subgroup. This four-atom phase is metallic but hosts three independent lattice vibration modes: a polar, an antipolar, and a Peierls mode. Condensing the Peierls mode required cell doubling that folds the BZ-boundary antipolar mode onto the $\Gamma$ of the experimentally observed ``eight atom" cell. The FE and AFE soft modes of the conventional PE phase ``eight atoms are therefore two \emph{distinct} order parameters of a common parent, brought to $\Gamma$ by band folding, rather than being two emergent phenomena from Peierls physics. We also realize that group theory forbids any trilinear coupling among the three order parameters, and we quantify all symmetry-allowed biquadratic couplings from first principles. A systematic comparison with isostructural TaOI$_2$, which shares the Peierls dimerization but shows no ferroelectricity, then isolates the origin of the polar instability in the Nb--O bonding channel and uncovers an unexpected sign reversal of the Peierls--polar coupling between the two chemistries.

\emph{The ``four atom unit cell"}: We begin by discussing the four-atom unit cell constructed in this work. Previous studies have described the PE phase of NbOI$_2$ as a monoclinic structure with $C2/m$ symmetry and eight atoms, or two formula units, per primitive cell \cite{rijnsdorp_crystal_1978}. In this structure, the Nb chains running along $\mathbf{a}$ are dimerized, reflecting the intrinsic Peierls distortion of the material. The phonon dispersion of this phase exhibits both FE and AFE instabilities at the center of its BZ. The presence of a $\Gamma$ AFE mode is unusual because it differs from the more familiar situation in conventional perovskite FEs, where antipolar instabilities generally occur at the BZ-boundary and therefore lead to a doubling of cells to condense in (see Fig.\ref{f1}d).

Applying a pseudosymmetry analysis to this structure, we find a translation that maps every atom onto the same-species partner up to systematic residuals that correspond exactly to the dimerization pattern. Symmetrizing these residuals away and halving the cell yields a one-formula-unit cell, that we refer to as the ``four-atoms cell" throughout this report. It also possesses the space group $C2/m$, in which Nb occupies Wyckoff position $2a$ with site symmetry $2/m$ --- which is located \emph{on an inversion center}. The observed ``8-atom" cell is an isomorphic subgroup of index 2 of this smaller cell. Construction and structural details are given in the Supplemental Material (SM). The ``4-atoms cell" is metallic by construction, which we confirm by electronic band structure calculation: with one $d^1$ Nb per cell, the chain band is half filled. Orbital-projected band structures (SM) show that the band at the Fermi level is a nearly pure Nb--Nb $d$ band (see Fig.\ref{f1}a). The band structure of the "8-atom" cell shows the expected semiconducting behavior (see Fig.\ref{f1}a), signs of the metal-insulator transition that follows from the Peierls mode condensation. The full electronic band of the two structures can be found in the SM.

\emph{Phonon modes and the folding argument}: The harmonic phonon spectrum of the NbOI$_2$ ``4-atoms cell" (See Fig.\ref{f1}c) shows three main modes (plus an interchain replica discussed in the SM): (i) a $\Gamma$ polar mode $P$ at $-1.63$~THz, driven by Nb displacement along the polar axis; (ii) a stable BZ-boundary antipolar mode $\eta$ at $\mathbf{A} = (0,0,\tfrac{1}{2})$, in which Nb atoms of adjacent cells displace in an antiphase manner along the same axis; and (iii) the Peierls mode $\delta$, also at $\mathbf{A}$, at $-7.24$~THz, corresponding to Nb--Nb chain dimerization.

Because $2\mathbf{A}$ is a valid reciprocal-lattice vector, condensation of
$\delta$ doubles the cell and folds the point $\mathbf{A}$ onto $\Gamma$. The two soft $\Gamma$ modes observed in the ``8-atom" PE phase are thus explained at once: the FE mode is the inherited ``4-atoms cell" phase mode $P$; and the AFE mode is the folded ``4-atoms cell" phase $\eta$. Their coexistence at $\Gamma$ is a consequence of the Peierls cell doubling, not a property of the Peierls distortion itself. 

Group theory analysis further supports this conclusion. Since Nb sits on an inversion center of the parent phase, every irreducible representation carrying a Nb displacement is odd under inversion. Using ISODISTORT we identify $P \sim \GMm$, $\eta \sim \Aone$, and $\delta \sim \Atwo$ (Miller--Love notation \cite{aroyo}), with Nb-site characters $A_u$, $A_u$, and $B_u$, respectively. The candidate trilinear invariant $\delta P \eta$ satisfies momentum conservation ($2\mathbf{A} + \Gamma \equiv \Gamma$) but fails the parity selection rule: $B_u \otimes A_u \otimes A_u = B_u \neq A_g$. No trilinear coupling among the three order parameters is therefore symmetry-allowed. The leading-order Landau energy thus involves purely biquadratic mode--mode couplings and is given by:
\begin{align}
E ={}& E_0 + \sum_{i}\Big[\tfrac{1}{2}a_i Q_i^2 + \tfrac{1}{4}b_i Q_i^4\Big]
      + \sum_{i<j}\lambda_{ij}\,Q_i^2 Q_j^2,
\label{eq:landau}
\end{align}
with $i \in \{P, \eta, \delta\}$. This expression has an immediate physical consequence: unlike in improper FEs, where a trilinear term forces a secondary polar order the moment the primary order parameter condenses \cite{benedek2011hybrid,glamazda2018soft}, here the Peierls distortion cannot \emph{induce} polarization in linear order in $P$.

\emph{First-principles Landau coefficients}: We extract all coefficients of Eq.~\eqref{eq:landau} from frozen-mode density-functional theory (DFT) calculations on the undimerized parent (methods are detailed  in SM). Mode amplitudes $Q$ are measured in \AA{} with unit-normalized displacement eigenvectors. Table~\ref{tab:landau} summarizes the results for both NbOI$_2$ and isostructural TaOI$_2$. Three features stand out.
\begin{table*}[t]
\caption{\label{tab:landau}%
Computed Landau coefficients of Eq.~\eqref{eq:landau}. Harmonic
stiffnesses $a_i$ in eV/\AA$^2$; quartic coefficients $b_i$ and
biquadratic couplings $\lambda_{ij}$ in eV/\AA$^4$. Amplitudes are in
\AA{} with unit-normalized eigenvectors. The TaOI$_2$ couplings $\lambda_{P\eta}$ and
$\lambda_{\delta\eta}$ are leading small-amplitude values.}
\begin{ruledtabular}
\begin{tabular}{lccccccccc}
 & $a_P$ & $a_\eta$ & $a_\delta$ &
 $b_P$ & $b_\eta$ & $b_\delta$ &
 $\lambda_{P\eta}$ & $\lambda_{\delta P}$ & $\lambda_{\delta\eta}$ \\
\colrule
NbOI$_2$ & $-2.6$ & $+0.6$ & $-34.9$ & $+1.0\times 10^3$ & $+0.5\times 10^3$ & $+6.2\times 10^3$ & $+1155$ & $+12$ & $-20$ \\
TaOI$_2$ & $+4.2$ & $+4.6$ & $-38.8$ & $+60$             & $+13$             & $+4.8\times 10^3$ & $+53$   & $-54$ & $-30$ \\
\end{tabular}
\end{ruledtabular}
\end{table*}
First, in NbOI$_2$ the Peierls stiffness $|a_\delta|$ exceeds the polar
one $|a_P|$ by more than an order of magnitude: $\delta$ is by far the
dominant instability of the parent phase, consistent with the fact that
the Peierls distortion condenses first and survives in all phases of the
material.

Second, the largest mode--mode coupling by far is not Peierls-mediated. It is the repulsive interaction between the polar and antipolar modes, $\lambda_{P\eta} = +1155$~eV/\AA$^4$, which exceeds the two $\delta$-mediated couplings by a factor of 50--100. 
The positive sign reveals that these two modes strongly compete: condensation of either one strongly disfavors the other, which naturally protects the observed single-domain-character FE ground state against FE/AFE admixture.

\begin{figure}[!ht]
	\centering
	\includegraphics[width=0.5\textwidth]{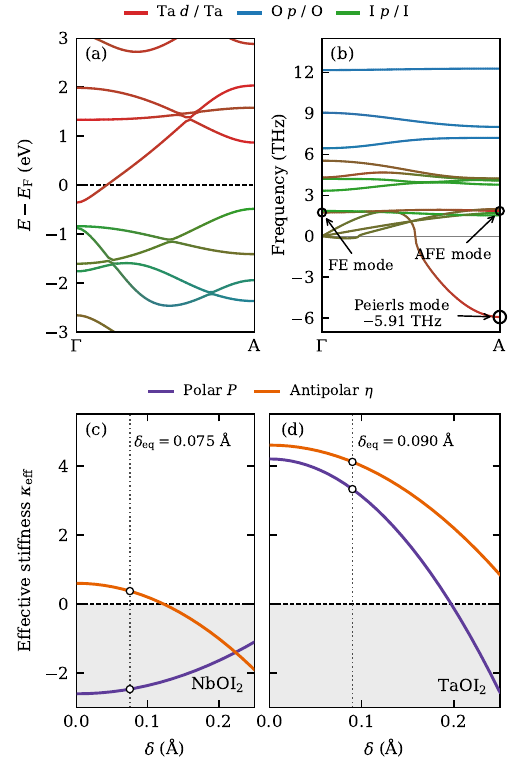}
	\caption{\label{f2} (a) Electronic band structure and (b) phonon dispersion of TaOI$_2$ along $\Gamma$-A in the 4-atom cell structure. Band colors show the orbital projections onto Ta d (red), O p (blue), and I p (green), and the dashed line marks the Fermi level. Phonon branches are colored by the atomic contributions of Ta, O, and I to the eigenvector. Like NbOI$_2$, the ``4 atom cell"  of TaOI$_2$ is metallic and has a strong Peierls instability at A (-5.91 THz). Unlike NbOI$_2$, however, its FE mode at $\Gamma$ (1.74 THz) and AFE mode at A (1.85 THz) are both stable. (c), (d) Effective stiffness of the polar (P, in blue) and antipolar ($\eta$, in red) modes as a function of the frozen Peierls amplitude $\delta$, $\kappa_{eff,i} = a_{i} + 2\lambda \delta_{i} \delta^{2}$, for (c) NbOI$_2$ and (d) TaOI$_2$. The dotted lines mark the equilibrium Peierls amplitude, $\delta_{eq} = (-a_{\delta}/b_{\delta})^{1/2}$, and the shaded region ($\kappa_{eff}<0$) indicates instability. At $\delta_{eq}$, the polar mode is unstable in NbOI$_2$ ($\kappa_{eff,P} = -2.46$), whereas both modes are stable in TaOI$_2$ ($\kappa_{eff,P}=3.33$). The polar mode in TaOI$_2$ softens only for $\delta \geq 0.20$ \AA, more than twice $\delta_{eq}$.}
\end{figure}
Third, the two $\delta$-mediated couplings have opposite signs in NbOI$_2$ implying the effective harmonic coefficients of the FE and AFE modes are:
\begin{equation}
a_P^{\rm eff}=a_P+2\lambda_{\delta P}\delta^2=-2.6+24\delta^2, 
\qquad
a_\eta^{\rm eff}=a_\eta+2\lambda_{\delta\eta}\delta^2=0.6-40\delta^2.
\end{equation}

Thus, increasing the Peierls amplitude hardens the polar mode, while strongly softening the antipolar mode and eventually driving it unstable (see Fig.\ref{f2}c). This is consistent with the phonon calculations, where the FE instability is hardened  upon dimerization whereas the antipolar mode, stable in the four-atom phase, becomes soft in the doubled eight-atom dimerized cell. The Peierls transition therefore acts, if anything, \emph{against} ferroelectricity contrary to what is commonly believed in the literature. Ferroelectricity survives in NbOI$_2$ because the equilibrium Peierls mode amplitude (see SM, Sec.~S4 and Fig.~\ref{fig:s4_bands}) is not enough to suppress its intrinsic instability. Interestingly, Fig.\ref{f2}c also suggests that the Peierls mode can be used to switch on and off ferroelectricity in NbOI$_2$. In other words, engineering a mechanism that enhances the Peierls distortion could provide an effective way to suppressing electrical polarization in NbOI$_2$. Optical illumination  might offer such a mechanism, as our previous study demonstrated that illumination enhances the Peierls distortion in NbOX$_2$ while simultaneously quenching its electrical polarization \cite{dansou2025photoinduced}. 

Now we turn to the following question: \emph{Why is TaOI$_2$ not ferroelectric?}
since TaOI$_2$ is isostructural to NbOI$_2$ and shares the $d^1$ chain filling and the Peierls dimerization. As shown in Fig.\ref{f2}a, its ``4-atoms cell", constructed identically, has an electronic structure nearly indistinguishable from that of NbOI$_2$: the same half-filled $d$--$d$ $\sigma$ chain band. For completness, the electronic band of the 8-atom cell of TaOI$_2$ is also shown in the SM.

The phonon spectrum of of TaOI$_2$ retains the deep Peierls instability ($-5.9$~THz) but shows \emph{no} unstable polar mode: the polar branch sits at $+1.7$~THz and the antipolar one at $+1.9$~THz. At the phenomenological level (Table~\ref{tab:landau}), the entire FE coefficient increases by several eV/\AA$^2$ namely $a_P$ from $-2.6$ to $+4.2$ and $a_\eta$ from $+0.6$ to $+4.6$; while $a_\delta$ changes by only $\sim$10\%. Strikingly, $\lambda_{\delta P}$ \emph{changes sign} from $+12$~eV/\AA$^4$ (competitive) in NbOI$_2$ to $-54$~eV/\AA$^4$ (cooperative) in TaOI$_2$. In NbOI$_2$ the chain dimerization works against polar order; in TaOI$_2$ it would promote it, while the polar mode is not stable to begin with. Yet no ferroelectricity is reported in TaOI$_2$. 

To understand why, we do again the above discussed effective stiffness analysis, but now for TaOI$_2$. From the coefficients in Table~\ref{tab:landau}, it can be established the following:

\begin{equation}
a_P^{\rm eff}=4.2-108\delta^2,
\qquad
a_\eta^{\rm eff}=4.6 -60\delta^2.
\end{equation}
These two effective coefficients, as function of the Peierls mode, are shown on Fig.\ref{f2}d. At the equilibrium Peierls mode, $a_P^{\rm eff}$ and $a_\eta^{\rm eff}$ are still all positive. This means that, even though the FE and Peierls mode are cooperative in TaOI$_2$, the equilibrium Peierls mode is not enough to induce ferroelectricity in this material. We also show in the SM the full potential well of the polar and the antipolar modes in NbOI$_2$ and TaOI$_2$ at $\delta=0$ and $\delta_{eq}$. 

From the above analyses, one can deduce that, while the Peierls physics is transferable between the two chemistries, the polar physics is not. Specifically, the analysis on TaOI$_2$ reveals that the nature of the interplay (i.e., competitive {\it versus} collaborative) between Peierls and polar orders is not a universal feature of the MOX$_2$ lattice. A microscopic analysis of this sign reversal, which we conjecture reflects the different orbital nature of the polar eigenvector in the two compounds (nearly pure Nb off-centering in NbOI$_2$ {\it versus} a strongly mixed Ta/I pattern in TaOI$_2$; SM), is an interesting subject for future work.

\emph{Discussion and conclusions.---}
In this study, we re-analyze the understanding of ferroelectricity in NbOI$_2$ around three statements. (i)~The polar instability is an independent order parameter of a metallic parent structure, not a by-product of the Peierls transition; symmetry forbids the trilinear coupling that an ``improper'' Peierls-driven mechanism would require, and the computed biquadratic coupling shows the Peierls distortion in fact mildly opposes polar order in NbOI$_2$. (ii)~The coexistence of FE and AFE soft modes at the zone center of the PE phase, which can not be explained in the Peierls-centric picture, follows immediately from the folding of the metallic parent BZ-boundary antipolar mode by the Peierls cell doubling, and their strong mutual competition ($\lambda_{P\eta}$) is the dominant anharmonic interaction of the material. (iii)~The comparison with TaOI$_2$ further decouples the polar mechanism from the Peierls physics: the Nb$\to$Ta substitution uniformly stabilizes the polar and antiplor modes while leaving the Peierls mode quantitatively intact.

Beyond NbOI$_2$, the  picture shown in this study, testing whether an observed ``PE reference'' is itself a condensed daughter of a smaller parent, and classifying couplings at the level of that parent, should apply broadly to low-dimensional materials in which Peierls, or other BZ-boundary modes coexist with polar mode, including the wider NbO$X_2$ family and related $d^1$ chain compounds. The sign reversal of the Peierls--polar coupling between Nb and Ta suggests a design principle worth exploring: in solid solutions Nb$_{1-x}$Ta$_x$OI$_2$, the effective $\lambda_{\delta P}$ should interpolate through zero, offering a chemistry knob over the interaction between dimerization and polarization. We leave this as an interesting future question to explore.

\begin{acknowledgments}
The authors acknowledge the Grant No. W911NF-25-1-0223 from the Army Research Office(ARO), the Vannevar Bush Faculty Fellowship (VBFF) Grant No.N00014-20-1-2834 from the Department of Defense, and the MonArk NSF Quantum Foundry supported by the National Science Foundation Q-AMASE-i Program under NSF Award No.
DMR-1906383. This research is also supported by the Arkansas High Performance Computing Center which is funded through multiple National Science Foundation grants and the Arkansas Economic Development Commission.  
\end{acknowledgments}

\bibliography{Refs}

@article{qi2021review,
  title={Review on recent developments in 2D ferroelectrics: theories and applications},
  author={Qi, Lu and Ruan, Shuangchen and Zeng, Yu-Jia},
  journal={Advanced Materials},
  volume={33},
  number={13},
  pages={2005098},
  year={2021},
  publisher={Wiley Online Library}
}

@article{jin2022rise,
  title={The rise of {2D} materials/ferroelectrics for next generation photonics and optoelectronics devices},
  author={Jin, Linghua and Wang, Huide and Cao, Rui and Khan, Karim and Tareen, Ayesha Khan and Wageh, Swelm and Al-Ghamdi, Ahmed A and Li, Shaojuan and Li, Dabing and Zhang, Ye and others},
  journal={Apl. Mat},
  volume={10},
  number={6},
  year={2022},
  publisher={AIP Publishing}
}

@article{abdelwahab_giant_2022,
	title = {Giant second-harmonic generation in ferroelectric {NbOI\ensuremath{_{2}}}},
	volume = {16},
	number = {9},
	journal = {Nat. Photon.},
	author = {Abdelwahab, Ibrahim and Tilmann, Benjamin and Wu, Yaze and Giovanni, David and Verzhbitskiy, Ivan and Zhu, Menglong and Berté, Rodrigo and Xuan, Fengyuan and Menezes, Leonardo de S. and Eda, Goki and Sum, Tze Chien and Quek, Su Ying and Maier, Stefan A. and Loh, Kian Ping},
	month = sep,
	year = {2022},
	pages = {644--650},
}

@article{jia_niobium_2019,
	title = {Niobium oxide dihalides {{NbOX\ensuremath{_{2}}}} : a new family of two-dimensional van der {Waals} layered materials with intrinsic ferroelectricity and antiferroelectricity},
	volume = {4},
	number = {5},
	journal = {Nanoscale Horizons},
	author = {Jia, Yinglu and Zhao, Min and Gou, Gaoyang and Cheng Zeng, Xiao and Li, Ju},
	year = {2019},
	pages = {1113--1123},
}

@article{ye_2023,
	title = {Manipulation of nonlinear optical responses in layered ferroelectric niobium oxide dihalides},
	volume = {14},
	issn = {2041-1723},
	number = {1},
	journal = {Nat. Commun.},
	author = {Ye, Liangting and Zhou, Wenju and Huang, Dajian and Jiang, Xiao and Guo, Qiangbing and Cao, Xinyu and Yan, Shaohua and Wang, Xinyu and Jia, Donghan and Jiang, Dequan and Wang, Yonggang and Wu, Xiaoqiang and Zhang, Xiao and Li, Yang and Lei, Hechang and Gou, Huiyang and Huang, Bing},
	year = {2023},
	pages = {5911},
}

@article{rijnsdorp_crystal_1978,
	title = {The crystal structure of niobium oxide diiodide {NbOI\ensuremath{_{2}}}},
	volume = {61},
	issn = {0022-5088},
	number = {1},
	urldate = {2024-05-20},
	journal = {Journal of the Less Common Metals},
	author = {Rijnsdorp, J and Jellinek, F},
	year = {1978},
	pages = {79--82}
}

@article{mohebpour_origin_2024,
	title = {Origin and properties of the flat band in monolayer {NbOCl\ensuremath{_{2}}}},
	volume = {110},
	number = {3},
	journal = {Phys. Rev. B},
	author = {Mohebpour, Mohammad Ali and Vishkayi, Sahar Izadi and Vitale, Valerio and Seriani, Nicola and Tagani, Meysam Bagheri},
	year = {2024},
	pages = {035429},
}

@article{ye_ferroelectric_2021,
	title = {Ferroelectric {Controlled} {Spin} {Texture} in {Two}-{Dimensional} {NbOI\ensuremath{_{2}}} {Monolayer}},
	volume = {38},
	number = {8},
	journal = {Chinese Phys. Lett.},
	author = {Ye, Qian and Shen, Yu-Hao and Duan, Chun-Gang},
	year = {2021},
	pages = {087702},
}

@article{zhang_giant_2024,
	title = {Giant electro-optic and elasto-optic effects in ferroelectric {NbOI\ensuremath{_{2}}}},
	volume = {110},
	number = {10},
	journal = {Phys. Rev. B},
	author = {Zhang, Zhenlong and Di, Xuehan and Paillard, Charles and Bellaiche, Laurent and Jiang, Zhijun},
	year = {2024},
	pages = {L100101},
}

@article{pbe,
  title={Generalized gradient approximation made simple},
  author={Perdew, John P and Burke, Kieron and Ernzerhof, Matthias},
  journal={Phys. Rev. Lett.},
  volume={77},
  number={18},
  pages={3865},
  year={1996},
  publisher={APS}
}

@article{choe2025,
  title={Observation of coherent ferrons},
  author={Choe, Jeongheon and Handa, Taketo and Huang, Chun-Ying and Liston, Andr{\'e} Koch and Cox, Jordan and Stensberg, Jonathan and Hong, Yongseok and Chica, Daniel G and Xu, Ding and Tay, Fuyang and others},
  journal={arXiv preprint arXiv:2505.22559},
  year={2025}
}

@article{togo20,
  title={First principles phonon calculations in materials science},
  author={Togo, Atsushi and Tanaka, Isao},
  journal={Scripta materialia},
  volume={108},
  pages={1--5},
  year={2015},
  publisher={Elsevier}
}

@article{kresse1996,
  title={Efficient iterative schemes for ab initio total-energy calculations using a plane-wave basis set},
  author={Kresse, Georg and Furthm{\"u}ller, J{\"u}rgen},
  journal={Physical review B},
  volume={54},
  number={16},
  pages={11169},
  year={1996},
  publisher={APS}
}

@article{kresse1999t,
  title={From ultrasoft pseudopotentials to the projector augmented-wave method},
  author={Kresse, Georg and Joubert, Daniel},
  journal={Physical review b},
  volume={59},
  number={3},
  pages={1758},
  year={1999},
  publisher={APS}
}

@article{dansou2025photoinduced,
  title={Photoinduced phase transitions and lattice deformation in two-dimensional {NbOX$_2$ (X= Cl, Br, I)}},
  author={Dansou, Carmel and Paillard, Charles and Bellaiche, Laurent},
  journal={Physical Review Materials},
  volume={9},
  number={11},
  pages={114001},
  year={2025},
  publisher={APS}
}

@article{iniguez2014first,
  title={First-principles study of the multimode antiferroelectric transition in {PbZrO$_3$}},
  author={{\'I}{\~n}iguez, Jorge and Stengel, Massimiliano and Prosandeev, Sergey and Bellaiche, L},
  journal={Physical Review B},
  volume={90},
  number={22},
  pages={220103},
  year={2014},
  publisher={APS}
}

@article{hlinka2014multiple,
  title={Multiple soft-mode vibrations of lead zirconate},
  author={Hlinka, J and Ostapchuk, T and Buixaderas, E and Kadlec, C and Kuzel, P and Gregora, I and Kroupa, J and Savinov, M and Klic, A and Drahokoupil, J and others},
  journal={Physical Review Letters},
  volume={112},
  number={19},
  pages={197601},
  year={2014},
  publisher={APS}
}

@article{benedek2011hybrid,
  title={Hybrid improper ferroelectricity: a mechanism for controllable polarization-magnetization coupling},
  author={Benedek, Nicole A and Fennie, Craig J},
  journal={Physical review letters},
  volume={106},
  number={10},
  pages={107204},
  year={2011},
  publisher={APS}
}

@article{glamazda2018soft,
  title={Soft tilt and rotational modes in the hybrid improper ferroelectric {Ca3Mn2O7}},
  author={Glamazda, A and Wulferding, D and Lemmens, P and Gao, B and Cheong, S-W and Choi, K-Y},
  journal={Physical Review B},
  volume={97},
  number={9},
  pages={094104},
  year={2018},
  publisher={APS}
}

@article{aroyo,
  title={Bilbao Crystallographic Server. II. Representations of crystallographic point groups and space groups},
  author={Aroyo, Mois I and Kirov, Asen and Capillas, Cesar and Perez-Mato, JM and Wondratschek, Hans},
  journal={Foundations of Crystallography},
  volume={62},
  number={2},
  pages={115--128},
  year={2006},
  publisher={International Union of Crystallography}
}

@article{li2025peierls,
  title={Peierls Distortion Induced Giant Linear Dichroism, Second-Harmonic Generation, and In-Plane Ferroelectricity in NbOBr2},
  author={Li, Zhuolun and Wang, Shaolong and Wang, Chong and Zhang, Xi and Yu, Zhipeng and Du, Shanzheng and Zeng, Zhongming and Dai, Yang and Zhang, Yuchao and Liu, Xiaochi and others},
  journal={Small},
  volume={21},
  number={2},
  pages={2407729},
  year={2025},
  publisher={Wiley Online Library}
}

@article{zhang2026electric,
  title={Electric-field control of coherent ferron oscillations},
  author={Zhang, Baolong and Duan, Ruihuan and Mishra, Sobhan Subhra and Jana, Sambhu and Kim, Jonghyeon and Tan Caiwei, Thomas and Tan, Yi Ji and Wang, Wenhao and Chen Ietro, Pang Teng and Liu, Zheng and others},
  journal={Nature Physics},
  pages={1--9},
  year={2026},
  publisher={Nature Publishing Group UK London}
}

@article{kresse1996efficiency,
  title={Efficiency of ab-initio total energy calculations for metals and semiconductors using a plane-wave basis set},
  author={Kresse, Georg and Furthm{\"u}ller, J{\"u}rgen},
  journal={Computational materials science},
  volume={6},
  number={1},
  pages={15--50},
  year={1996},
  publisher={Elsevier}
}

@article{grimme2010consistent,
  title={A consistent and accurate ab initio parametrization of density functional dispersion correction (DFT-D) for the 94 elements H-Pu},
  author={Grimme, Stefan and Antony, Jens and Ehrlich, Stephan and Krieg, Helge},
  journal={The Journal of chemical physics},
  volume={132},
  number={15},
  year={2010},
  publisher={AIP Publishing}
}

@article{ding2025strain,
  title={Strain-Induced Ferroelectricity and Second-Harmonic Generation in a {TaOI2} Monolayer},
  author={Ding, Yi-min and Wang, Qiang and Jiang, Min and Zhou, Liujiang and Li, Youyong},
  journal={The Journal of Physical Chemistry Letters},
  volume={16},
  number={46},
  pages={12061--12067},
  year={2025},
  publisher={ACS Publications}
}

@article{ruck1995taoi2,
  title={{TaOI2}: Eine zentrosymmetrische Variante der {NbOI2}-Struktur},
  author={Ruck, Michael},
  journal={Crystal Structure Communications},
  volume={51},
  number={10},
  pages={1960--1962},
  year={1995},
  publisher={International Union of Crystallography}
}

@article{tang2024third,
  title={Third harmonic generation in thin {NbOI2} and {TaOI2}},
  author={Tang, Tianhong and Hu, Deng and Lin, Di and Yang, Liu and Shen, Ziling and Yang, Wenchen and Liu, Haiyang and Li, Hanting and Fan, Xiaoyue and Wang, Zhiwei and others},
  journal={Nanomaterials},
  volume={14},
  number={5},
  pages={412},
  year={2024},
  publisher={MDPI}
}

@misc{isotropy,
  author = {Stokes, H. T. and Hatch, D. M. and Campbell, B. J.},
  title = {{ISOTROPY} Software Suite},
  howpublished = {\url{https://iso.byu.edu}},
  note = {Accessed: 2026-10-01}
}

\clearpage
\onecolumngrid
\begin{center}
\textbf{\large Supplemental Material for ``Ferroelectricity and Peierls transition in MOI$_2$ (M=Ta, Nb): a first-principle analysis''}\\[0.5em]
Carmel Dansou, Charles Paillard, and Laurent Bellaiche
\end{center}
\vspace{1em}
\twocolumngrid

\setcounter{equation}{0}
\setcounter{figure}{0}
\setcounter{table}{0}
\setcounter{section}{0}
\renewcommand{\theequation}{S\arabic{equation}}
\renewcommand{\thefigure}{S\arabic{figure}}
\renewcommand{\thetable}{S\arabic{table}}
\makeatletter
\renewcommand{\theHequation}{S\arabic{equation}}
\renewcommand{\theHfigure}{S\arabic{figure}}
\renewcommand{\theHtable}{S\arabic{table}}
\makeatother

\section*{S1. Pseudosymmetry construction of the parent phase, method details and structural data}

\subsection*{S1.A. Construction of the one-formula-unit parent}

The paraelectric reference structure conventionally used for \NbOI\ is the centrosymmetric, Peierls-dimerized $C2/m$ phase containing two formula units in its primitive cell.  In this structure, the Nb-Nb chain is already dimerized, so the translation-symmetry breaking associated with the Peierls order parameter is built into the reference cell.  To obtain the undimerized parent used in the present study, we searched the eight-atom structure for approximate translations that map each atom onto a same-species partner.

In the primitive setting used for this construction, the approximate translation $\mathbf{t}=(0,0,1/2)$ maps the two halves of the eight-atom cell onto one another.  For the input structure used in the pseudosymmetry analysis, the residual mismatch under this translation is approximately $0.64$~\angstrom\ on Nb, $0.56$~\angstrom\ on O, and $0.41$~\angstrom\ on I. We remove these residuals by pairwise symmetrization. Explicitly, if atoms $j$ and $j'$ are related by translation $\mathbf{t}=(0,0,1/2)$, their fractional coordinates are first brought into the same periodic image and replaced by the pair average compatible with $\mathbf{t}$.  The primitive vector parallel to the Peierls modulation direction is then halved.  The resulting one-formula-unit primitive cell contains four atoms and retains space-group type $C2/m$ (No.~12). The primitive cell contains one formula unit with Nb lying exactly on an inversion center. From symmetry perspective, the eight-atom paraelectric phase and the four-atom parent are related by an isomorphic subgroup relation of index two: condensation of the Peierls distortion doubles the primitive translation along the chain and restores the eight-atom periodicity.  In reciprocal space,  this doubling folds the parent phase BZ boundary point $\mathbf{A}=(0,0,1/2)$ onto $\Gamma$ of the doubled cell.  This folding is what brought the polar and the antipolar modes at $\Gamma$ of the eight-atom phase that is discussed in the main text.

The distortion connecting the symmetrized parent to the dimerized eight-atom paraelectric structure is dominated by opposite displacements of neighboring Nb atoms along the chain.  In the structural comparison used during the construction, the individual atomic displacements are approximately $\pm0.32$~\angstrom\ on Nb and $\pm0.28$~\angstrom\ on O along the chain direction, while I atoms undergo transverse displacements of approximately $\pm0.21$~\angstrom.  These Cartesian displacements should not be confused with the scalar normal-coordinate amplitude $Q_\delta$ used in the frozen-mode calculations, because $Q_\delta$ is defined with a normalized eigenvector [Sec.~S2.B].

\subsection*{S1.B. Methods}

First-principles calculations were carried out with the Vienna \textit{ab initio} Simulation Package (VASP) \cite{kresse1996,kresse1996efficiency}.  Exchange and correlation were treated within the Perdew--Burke--Ernzerhof generalized-gradient approximation \cite{pbe}, and the projector-augmented-wave (PAW) method was used \cite{kresse1999t}.  A plane-wave kinetic-energy cutoff of 600~eV was employed in all our calculations. For structural relaxations, we used $\texttt{EDIFF}=10^{-6}$~eV and a force criterion of $10^{-3}$~eV/\angstrom. For band structure, phonon and coupling constant calculations, the electronic self-consistency threshold was $\texttt{EDIFF}=10^{-8}$~eV. The Brillouin zone was sampled with a $8 \times 8\times 8$ $\Gamma$-centered mesh for the four-atom primitive cell and the corresponding commensurate meshes for doubled supercells. vdW interaction as implemented using the DFT-D3 method of Grimme with zero-damping function \cite{grimme2010consistent}. For static frozen-mode energies, the ionic coordinates and lattice vectors were held fixed and only the electronic degrees of freedom were converged.

\subsection*{S1.C. Structural tables}

Tables~\ref{tab:s1_lattice} and \ref{tab:s1_coords} list the exact structures used in our calculations. 

\begin{table}[!ht]
\caption{Lattice vectors of the four-atom cell and the eight-atom Peierls-dimerized paraelectric cell.  All entries are in \angstrom.}
\label{tab:s1_lattice}
\begin{ruledtabular}
\begin{tabular}{c c c c c}
Structure & vector & $x$ & $y$ & $z$ \\
\hline
4-atom PE & $\mathbf a_1$ & $ 7.6679$ & $ 1.9396$ & $ 0.0000$ \\
                  & $\mathbf a_2$ & $-7.6679$ & $ 1.9396$ & $ 0.0000$ \\
                  & $\mathbf a_3$ & $-0.8454$ & $ 0.0000$ & $ 3.6769$ \\
8-atom PE         & $\mathbf a_1$ & $-0.0992$ & $ 1.9396$ & $ 7.7495$ \\
                  & $\mathbf a_2$ & $-0.0992$ & $-1.9396$ & $ 7.7495$ \\
                  & $\mathbf a_3$ & $ 7.3006$ & $ 0.0000$ & $-1.9070$ \\
\end{tabular}
\end{ruledtabular}
\end{table}

\begin{table*}[!ht]
\caption{Fractional coordinates of the parent phase and the dimerized eight atom cell used in our calculations.}
\label{tab:s1_coords}
\begin{ruledtabular}
\begin{tabular}{ c c c c c}
Structure & atom & $x$ & $y$ & $z$ \\
\hline
4-atom PE & Nb & $0.0000$ & $0.0000$ & $0.0000$ \\
                  & O  &  $0.5000$ & $0.5000$ & $0.0000$ \\
                  & I$_1$ & $0.8603$ & $0.1397$ & $0.3740$ \\
                  & I$_2$ & $0.1397$ & $0.8603$ & $0.6260$ \\
\hline
8-atom PE & Nb$_1$ & $0.0006$ & $0.0006$ & $0.2919$ \\
          & Nb$_2$ &$0.9994$ & $0.9994$ & $0.7081$ \\
          & O$_1$  & $0.5011$ & $0.5011$ & $0.2859$ \\
          & O$_2$  & $0.4989$ & $0.4989$ & $0.7141$ \\
          & I$_1$  & $0.1266$ & $0.1266$ & $0.0676$ \\
          & I$_2$  & $0.8734$ & $0.8734$ & $0.9324$ \\
          & I$_3$  & $0.1529$ & $0.1529$ & $0.5859$ \\
          & I$_4$  & $0.8471$ & $0.8471$ & $0.4141$ \\
\end{tabular}
\end{ruledtabular}
\end{table*}

The four-atom structure is used here as a crystallographic and Landau reference, not as a thermodynamically accessible phase.  It contains a half-filled metallic chain band and a very strong Peierls instability, so it immediately lowers its energy by allowing the doubled periodicity.  The existence of this unstable parent nevertheless allows the polar, antipolar, and Peierls coordinates to be defined before dimerization and therefore makes it easier to define their symmetry relations.

\section*{S2. Phonon calculations and eigen modes mapping}

\subsection*{S2.A. Full phonon dispersion}
Harmonic force constants were obtained with the finite-displacement method as implemented in Phonopy \cite{togo20}.  For the four-atom structure, the full-dispersion calculations used a $2\times2\times4$ supercell while a $2\times2\times2$ supercell was used for the 8-atom structure.

\begin{figure*}[t]
\centering
\includegraphics[width=0.88\textwidth]{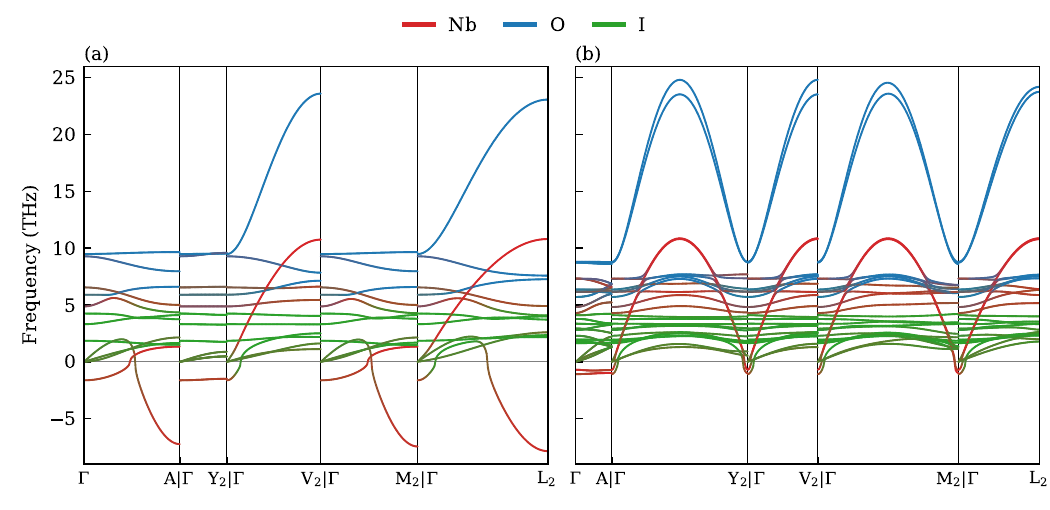}
\caption{Phonon dispersions of NbOI$_2$ in (a) the 4-atom and (b) the 8-atom cell, colored by the Nb (red), O (blue), and I (green) contributions to the eigenvector displacements. Imaginary frequencies are plotted as negative values. High-symmetry labels refer to the reciprocal basis of each cell.}
\label{fig:s2}
\end{figure*}

In the main text, we focus on the polar, antipolar, and Peierls coordinates that control the relation between the four- and eight-atom structures.  Figure~\ref{fig:s2} shows the complete phonon dispersion of the four-atom \NbOI\ -- which was obtained from finite-displacement force constants. The three order parameters emphasized in the Letter are the zone-center polar coordinate $P$, the boundary antipolar coordinate $\eta$ at $A$, and the boundary Peierls coordinate $\delta$ at $A$.  A fourth low-energy branch occurs at the in-plane boundary point $\mathbf{Y}_2=(-1/2,1/2,0)$, and is denoted as $\gamma$.  Its eigenvector consists predominantly of metal displacements along the polar direction, with neighboring chains moving out of phase. It is therefore the interchain-antipolar counterpart of the in-phase $\Gamma$-point polar mode.

\subsection*{S2.B. Phonons and eigenvector normalization}

For frozen-mode calculations, we define a real-space displacement vector $\hat{\mathbf u}^{(i)}$ for mode $i$ and normalize it in Cartesian displacement space as
\begin{equation}
 \sum_{\kappa\alpha}\left|\hat u^{(i)}_{\kappa\alpha}\right|^2=1.
 \label{eq:unit_normalization}
\end{equation}
The scalar mode amplitude $Q_i$, expressed in \angstrom, is then defined through
\begin{equation}
 \Delta R_{\kappa\alpha}=Q_i\hat u^{(i)}_{\kappa\alpha}.
 \label{eq:qdef}
\end{equation}
For a zone-boundary mode, the phase factor associated with its wave vector is included when the eigenvector is expanded into the commensurate real-space supercell.  The overall sign of an eigenvector is arbitrary; all energy fits therefore use symmetric positive and negative amplitudes.

\subsection*{S2.C. Phonon mode decomposition.}
Phonon eigenvectors $\mathbf{e}_\kappa(\mathbf{q}\nu)$ of the 4-atom phase cell were
taken from the \textsc{phonopy} output at $\Gamma$ (polar mode) and at A (antipolar mode).
These are eigenvectors of the dynamical matrix, i.e.\ mass-weighted and normalized as
$\sum_\kappa |\mathbf{e}_\kappa|^2 = 1$, where $\kappa$ runs over the atoms in the cell.
We quantify the contribution of each atom in two ways. The first is the eigenvector weight
$w_\kappa = |\mathbf{e}_\kappa|^2$. The second is the displacement weight
\begin{equation}
  \label{eq:disp_weight}
  \tilde{w}_\kappa = \frac{|\mathbf{u}_\kappa|^2}{\sum_{\kappa'} |\mathbf{u}_{\kappa'}|^2},
  \qquad \mathbf{u}_\kappa = \frac{\mathbf{e}_\kappa}{\sqrt{m_\kappa}},
\end{equation}
where $\mathbf{u}_\kappa$ is the atomic displacement and $m_\kappa$ the atomic mass.
Both weights are summed over symmetry-equivalent atoms to give species contributions.
$\tilde{w}_\kappa$ reflects the actual amplitude of atomic motion, whereas $w_\kappa$
gives more importance to heavy atoms. The two therefore differ noticeably when atoms of
very different masses are involved; for example, O takes 2\% of the eigenvector weight but
22\% of the displacement weight in the antipolar mode of TaOI$_2$.
At A, the complex eigenvector is multiplied by a global phase so that its largest
component is real; this choice does not affect the weights.
The modes of the two materials were matched by the eigenvector overlap
$|\langle \mathbf{e}^{\mathrm{Nb}} | \mathbf{e}^{\mathrm{Ta}} \rangle|$ at the same
$\mathbf{q}$, which gives 0.98 for the polar mode and 0.99 for the antipolar mode.

\begin{table}[!ht]
\caption{Atomic contributions to the FE mode at $\Gamma$ and the AFE mode at A in the 4-atom cell of NbOI$_2$ and TaOI$_2$. For each atom we give the eigenvector weight $w=|\mathbf{e}_\kappa|^2$ and, in parentheses, the displacement weight $\tilde{w}$ defined in Eq.~\eqref{eq:disp_weight}. Contributions of the two I atoms are summed. $M$ = Nb or Ta.}
\label{tab:mode_weights}
\begin{ruledtabular}
\begin{tabular}{llcccc}
Material & Mode & $\nu$ (THz) & $M$ & O & I \\
\hline
NbOI$_2$ & FE ($\Gamma$) & $-1.63$ & 0.74 (0.79) & 0.00 (0.01) & 0.26 (0.21) \\
         & AFE (A)       & $1.33$  & 0.98 (0.94) & 0.01 (0.05) & 0.02 (0.01) \\
TaOI$_2$ & FE ($\Gamma$) & $1.74$  & 0.57 (0.48) & 0.00 (0.02) & 0.42 (0.50) \\
         & AFE (A)       & $1.85$  & 0.98 (0.78) & 0.02 (0.22) & 0.00 (0.00) \\
\end{tabular}
\end{ruledtabular}
\end{table}

\begin{table}[!ht]
\caption{Eigenvector components $e_{\kappa,y}$ of the FE and AFE modes of NbOI$_2$ and TaOI$_2$ (4-atom cell). Only the Cartesian $y$ component, along the $M$--O chain, is nonzero; the $x$ and $z$ components vanish by symmetry.}
\label{tab:eigenvectors}
\begin{ruledtabular}
\begin{tabular}{llccccc}
Material & Mode & $\nu$ (THz) & $M$ & O & I$_1$ & I$_2$ \\
\hline
NbOI$_2$ & FE ($\Gamma$) & $-1.63$ & $+0.859$ & $-0.032$ & $-0.362$ & $-0.362$ \\
         & AFE (A)       & $1.33$  & $+0.988$ & $+0.089$ & $+0.034$ & $+0.034$ \\
TaOI$_2$ & FE ($\Gamma$) & $1.74$  & $+0.758$ & $+0.046$ & $-0.460$ & $-0.460$ \\
         & AFE (A)       & $1.85$  & $+0.988$ & $+0.156$ & $-0.003$ & $-0.003$ \\
\end{tabular}
\end{ruledtabular}
\end{table}

\section*{S3. Orbital-projected electronic structures and magnetic-instability checks}

\subsection*{S3.A. Orbital-projected bands structures}

\begin{figure*}[!ht]
\centering
\includegraphics[width =0.88\textwidth]{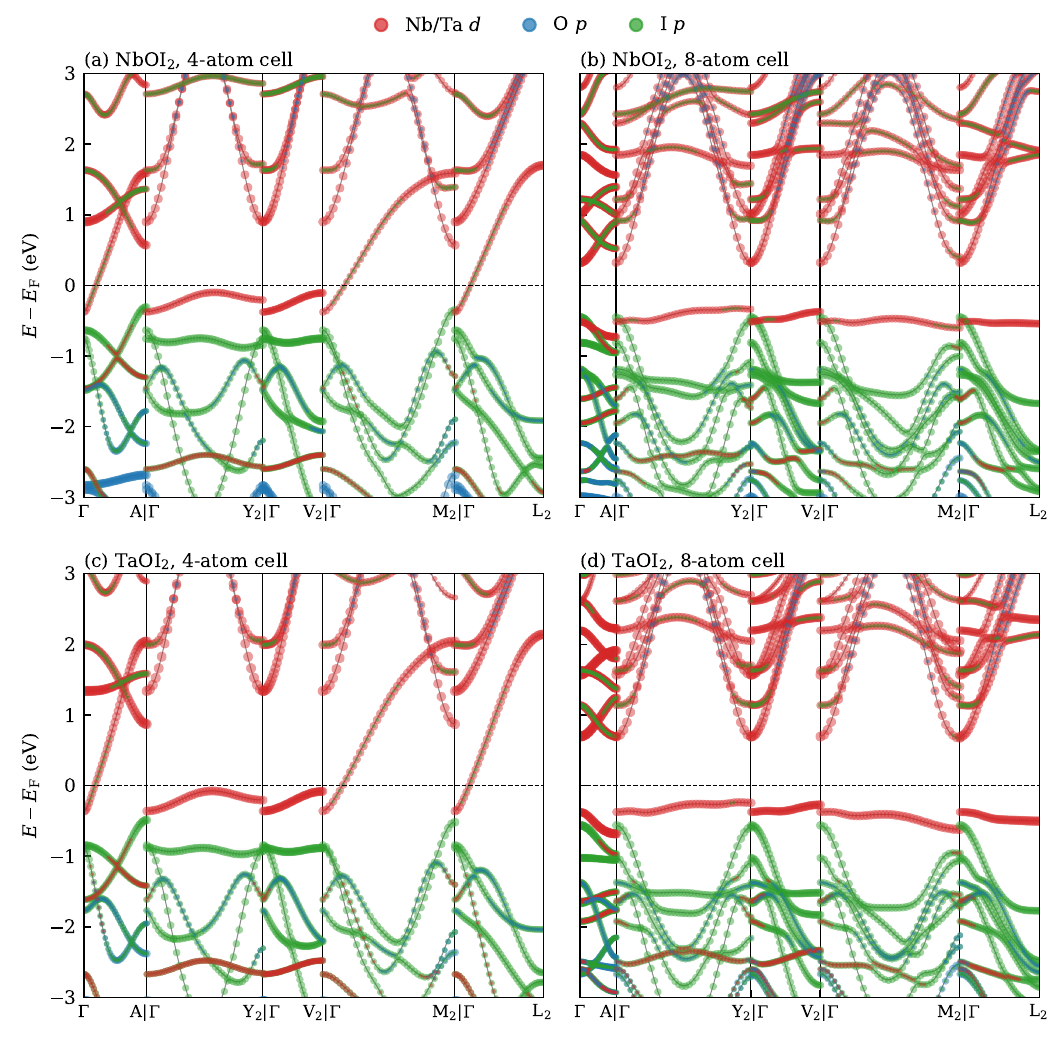}
\caption{Orbital-projected band structures of (a) NbOI$_2$ in the 4-atom cell, (b) NbOI$_2$ in the 8-atom cell, (c) TaOI$_2$ in the 4-atom cell, and (d) TaOI$_2$ in the 8-atom cell. Circle sizes are proportional to the Nb/Ta d (red), O p (blue), and I p (green) projections, and the dashed line marks the Fermi level. Both 4-atom cells are metallic. The 8-atom cells open gaps of 0.65 eV (NbOI$_2$) and 0.90 eV (TaOI$_2$).}
\label{fig:s3}
\end{figure*}

The four-atom parent phases of \NbOI\ and \TaOI\ are metallic by construction: each contains one $d^1$ metal ion per primitive cell and therefore an odd number of valence electrons for the PAW choices used in the calculations.  Figure \ref{fig:s3} shows the orbital-projected band structures of both  along the same $C2/m$ path, with metal $d$, iodine $p$, and oxygen $p$ orbitals resolved.  For \NbOI, the half-filled band crossing $E_F$ is approximately 88\% Nb $d$, 11\% I $p$, and 0.2\% O $p$. The Fermi-level state are predominantly metal $d$--$d$ band. The parent phase of \TaOI\ exhibits the same qualitative electronic structure: a half-filled, strongly chain-dispersive Ta $d$ band crosses the Fermi energy, with a small O contribution.

In the Peierls condensed phase, the two materials show semiconducting band with their band structure being qualitatively the same. One can see the nearly flat band around the Fermi level, characteristic of these materials. In the semiconducting state, the bands near the Fermi level are also predominantly metal-metal $d$--$d$ states as in the parent metallic phase.

\section*{S4. Symmetry invariants and frozen-mode fitting procedure}

\subsection*{S4.A. Irreducible representations and selection rules}

We classified the relevant normal modes at $\Gamma$, and $A$ using group irreducible representations of the four-atom $C2/m$ phase.  The final labels used in the Letter are summarized in Table~\ref{tab:s5_irreps}.  The essential result is that the three primary order parameters $P$, $\eta$, and $\delta$ are distinct symmetry modes of the same undimerized parent and that the candidate trilinear invariant $\delta P\eta$ is forbidden.

\begin{table}[!ht]
\caption{Mode labels used in the main text.  The Miller--Love labels are copied directly from the production ISODISTORT/irrep analysis \cite{isotropy}.}
\label{tab:s5_irreps}
\begin{ruledtabular}
\begin{tabular}{c c c c}
mode & symbol & $\mathbf q$ & irrep \\
\hline
polar & $P$ & $\Gamma$ & $\Gamma_1^-$ \\
antipolar & $\eta$ & $A$ & $A_1^-$ \\
Peierls & $\delta$ & $A$ & $A_2^-$ \\
\end{tabular}
\end{ruledtabular}
\end{table}

A monomial in the Landau expansion is allowed only if (i) the sum of its wave vectors is a reciprocal-lattice vector and (ii) the direct product of its irreducible representations contains the identity representation at $\Gamma$.  All bilinear couplings between $P$ and the BZ boundary modes are forbidden by wave-vector conservation.  The bilinear $\eta\delta$ is also forbidden because, although both modes occur at $A$, their irreducible representations are different and their product does not transform as the identity.

The trilinear term $\delta P\eta$ satisfies momentum conservation because
\begin{equation}
 A+\Gamma+A=2A\equiv G,
\end{equation}
but it fails the representation/parity condition.  With the final irrep used in the main text, the three modes are inversion-odd at their respective invariant wave vectors, so the product of the three mode amplitudes is odd rather than totally symmetric.  Hence
 $\delta P\eta\quad$\text{is forbidden}.
This is the symmetry statement that excludes a lowest-order improper-ferroelectric mechanism in which the Peierls order parameter would linearly induce $P$ through the antipolar mode.

The lowest-order inter-mode terms are therefore biquadratic.  Restricting first to the three order parameters discussed in the Letter,
\begin{align}
E(P,\eta,\delta)=E_0
&+\frac12a_PP^2+\frac14b_PP^4
+\frac12a_\eta\eta^2+\frac14b_\eta\eta^4 \nonumber\\
&+\frac12a_\delta\delta^2+\frac14b_\delta\delta^4 \nonumber
+\lambda_{P\eta}P^2\eta^2 +\lambda_{\delta P}\delta^2P^2 \\
&+\lambda_{\delta\eta}\delta^2\eta^2+\cdots .
\label{eq:landau3}
\end{align}
The same symmetry logic allows $P^2\gamma^2$, $\eta^2\gamma^2$, and $\delta^2\gamma^2$.  A sixth-order invariant $\delta^2P^2\eta^2$ is also allowed and provides one route by which a finite Peierls amplitude can renormalize the effective $P$--$\eta$ competition beyond the leading small-amplitude model. It is not considered in this analysis.[{\bf Carmel need to insert a figure of mode vs energy from DFT and Landau free energy here}]

\subsection*{S4.B. Two-dimensional frozen-mode scans}

The coupling coefficients were extracted from single-point DFT calculations in a commensurate $1\times1\times2$ supercell of the undimerized phase, which contains eight atoms and can represent both $\Gamma$ and $A$ modes simultaneously.  For each pair $(X,Y)\in\{(P,\eta),(\delta,P),(\delta,\eta)\}$, the atomic coordinates were displaced according to the following equation:
\begin{equation}
 \mathbf R(Q_X,Q_Y)=\mathbf R_0+Q_X\hat{\mathbf u}_X+Q_Y\hat{\mathbf u}_Y,
\end{equation}
with no ionic or lattice relaxation.  The amplitude grid used was:
\begin{equation}
 Q_X,Q_Y\in\{\-0.04, 0.04\}\ \angstrom.
\end{equation}

\begin{figure*}[!ht]
\centering
\includegraphics[width =0.88\textwidth]{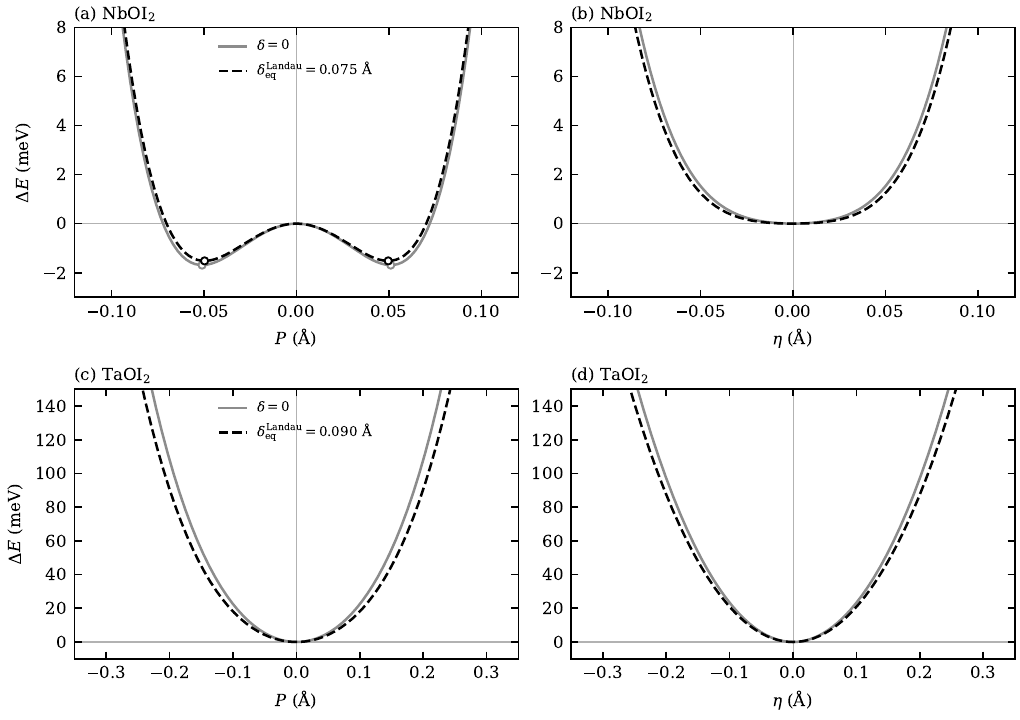}
\caption{Landau energy landscapes of the polar (P) and antipolar ($\eta$) modes of (a), (b) NbOI$_2$ and (c), (d) TaOI$_2$, at fixed Peierls amplitude $\delta$. At fixed $\delta$, each mode obeys $\Delta E(Q) = 1/2 \kappa_{eff}Q^2 + 1/4b Q^4$. The coefficients are listed in Table~\ref{tab:landau} of the main text. Curves are shown for the undistorted cell ($\delta$ = 0, grey solid) and for the equilibrium Peierls amplitude of the model, $\delta_{eq} = (-a_{\delta}/b_{\delta})^{1/2}$ (black dashed): 0.075 \AA~ for NbOI$_2$ and 0.090 \AA~  for TaOI$_2$. Open circles mark the energy minima. In NbOI$_2$, the polar mode remains a double well at $\delta_{eq}$, while the antipolar mode remains stable. In TaOI$_2$, condensing the Peierls mode softens the polar and antipolar modes slightly but not enough to make them unstable. }
\label{fig:s4_bands}
\end{figure*}

For each pair, the energies were fitted to
\begin{align}
E(Q_X,Q_Y)=E_0&+\frac12a_XQ_X^2+\frac14b_XQ_X^4
+\frac12a_YQ_Y^2+\frac14b_YQ_Y^4 \nonumber\\
&+\lambda_{XY}Q_X^2Q_Y^2,
\label{eq:2d_fit}
\end{align}
using the known least squares method.  Odd cross terms were excluded by symmetry.  On Fig. \ref{fig:s4_bands}, we show plot of the Landau energy as function of the polar and antipolar modes for the two materials at some fixed value of the Peierls mode. 

\end{document}